 \documentclass[preprint2]{aastex} 

\slugcomment{}

\shorttitle{The prediction method of similar cycles}
  \shortauthors{Z. L. Du \& H. N. Wang}

\begin{document}
\title{The prediction method of similar cycles}

\author{Z. L. Du and H. N. Wang\altaffilmark{}}
\affil{Key Laboratory of Solar Activity, National Astronomical
Observatories, Chinese Academy of Sciences, Beijing 100012, China}
\email{zldu@nao.cas.cn}

\begin{abstract}
The concept of degree of similarity ($\eta$) is proposed to
quantitatively describe the similarity of a parameter (e.g. the
maximum amplitude $R_{\mathrm{max}}$) of a solar cycle relative to
a referenced one, and the prediction method of similar cycles is
further developed. For two parameters, the solar minimum
($R_{\mathrm{min}}$) and rising rate ($\beta_{\mathrm{a}}$), which
can be directly measured a few months after the minimum, a
synthesis degree of similarity ($\eta_{\rm s}$) is defined as the
weighted-average of the $\eta$ values around $R_{\mathrm{min}}$
and $\beta_{\mathrm{a}}$ with the weights given by the
coefficients of determination of $R_{\mathrm{max}}$ with
$R_{\mathrm{min}}$ and $\beta_{\rm a}$, respectively. The monthly
values of the whole referenced cycle can be predicted by averaging
the corresponding values in the most similar cycles with the
weights given by the $\eta_{\rm s}$ values. Cycles 14 and 10 are
found to be the two most similar cycles of Cycle 24. As an
application, Cycle 24 is predicted to peak around January 2013
$\pm$ 8 (months) with a size of about $R_{\mathrm{max}}=84\pm 17$
and to end around September 2019.
\end{abstract}


\section{Introduction}           
\label{sect:intro}

Solar activity prediction is important for both space weather and
solar physics. Since solar activity is the driver of various
phenomena in the near-Earth environment, knowing the future level
of solar activity in advance can reduce some of the hazards for
high-tech equipment on which our modern society dependents.
Successful predictions could provide some constraints on solar
dynamo models \citep{Cameron08,Pesnell08,WangHN09,Guo10,Du11a}.

Various techniques have been used in the past to predict solar
activity, of which some were purely statistical and others were
related to physics \citep{Kane07a,Du308,Pesnell08,Messerotti09}.
Geomagnetic precursor methods have attracted more attention due to
successes in Cycles 20-22 \citep{Ohl66,Kane07a,Du309a,Du211a}.
They are based on a solar dynamo concept that the geomagnetic
activity during the declining phase of the preceding cycle or at
the minimum provides an approximate measure of the poloidal solar
magnetic field that generates the toroidal field for the next
cycle \citep{Schatten78}. Solar dynamo models have recently been
attempted, but they are as yet immature to predict the associated
solar activity \citep{Dikpati06,Tobias06,Pesnell08}.

A prominent feature in solar activity is the so-called Waldmeier
effect, where stronger cycles tend to rise faster
\citep{Waldmeier39,Hathaway02,Du309b}. This effect implies that
the magnetic energy in a solar cycle has a tendency of stability
and that stronger cycles need less time to release their energies
\citep{Du06a}. Correspondingly, the maximum amplitude
($R_{\mathrm{max}}$) of a solar cycle is well correlated with the
growth rate of activity in the early phase \citep{Cameron08}.

As a new solar cycle is ongoing, its shape can be well described
by simple functions containing a few parameters
\citep{Elling92,Hathaway94,Li99,Volobuev09,Du11b}. Before the
arrival of the peak of a solar cycle, only two parameters are
known: the preceding minimum
($R_{\mathrm{min}}$) and the 
rising rate ($\beta_{\mathrm{a}}$), defined as the ratio of the
increment of the activity ($R_{\mathrm{z}}$) and the elapsed time
duration. The two parameters can be taken as indicators for the
subsequent amplitude ($R_{\mathrm{max}}$).

Solar cycles that have approximatively the same $R_{\mathrm{max}}$
tend to have similar shapes, and these cycles are therefore called
`similar cycles' \citep{Waldmeier36,Gleissberg71,WangJL02a}.
\cite{Gleissberg71} used the concept that similar cycles tend to
have similar cycle lengths and similar decline times
\citep{Gleissberg73} to estimate the epochs of the start and end
of Cycle 21 by averaging the cycle lengths and decline times of
similar cycles, respectively. \cite{WangJL97} developed the
concept of similar cycles and used it to predict
$R_{\mathrm{max}}$ \citep{WangJL09} and monthly values of
$R_{\mathrm{z}}$ \citep{WangJL02a,Miao08}. The predicted value of
$R_{\mathrm{z}}$ for month $i$ is taken as the average of the
corresponding values of similar cycles for the same month from the
start of the cycles, and the standard deviation of these values is
taken as the prediction error for the same month.

The data and parameters used in this study are shown in Section
\ref{sec:data}. Some of the correlations between these parameters
are analyzed in Section \ref{sec:Correlation}, and in Section
\ref{sec:Similar}, we employ the two parameters of
$R_{\mathrm{min}}$ and $\beta_{\mathrm{a}}$ to further develop the
concept of a similar cycle and its application in prediction. In
Section \ref{subsec:similarity}, a quantity, the ``degree of
similarity ($\eta$)", is proposed to describe the similarity of a
parameter in past cycles relative to a referenced one (the
predictor for a cycle to be predicted). For two predictors
($R_{\mathrm{min}}$, $\beta_{\mathrm{a}}$), in
Section~\ref{subsec:Rm}, a synthesis degree of similarity
($\eta_{\mathrm{s}}$) is defined as the weighted-average of the
corresponding $\eta$ values ($\eta_{\mathrm{R}}$,
$\eta_{\mathrm{\beta}}$) with the weights given by the
coefficients of determination of $R_{\mathrm{max}}$ with
$R_{\mathrm{min}}$ and $\beta_{\rm a}$ ($r^2_{\rm R}$, $r^2_{\rm
\beta}$), respectively. Then, the $R_{\mathrm{max}}$ value for the
current cycle (24) can be predicted as the weighted-average of the
$R_{\mathrm{max}}$ values of the five most similar cycles with the
weights given by $\eta_{\mathrm{s}}$. The same technique is used
in Section~\ref{subsec:shape} for each month from the start of the
similar cycles to obtain the monthly values (shape) of Cycle 24.
The predictive power of the similar-cycle prediction method is
tested on different months from the start of Cycle 24 in
Section~\ref{subsec:m}. The results are briefly discussed and
summarized in Section \ref{sec:Discussions}.

\section{Data and cycle parameters} \label{sec:data}

The data used in this study are the smoothed monthly mean
Z\"{u}rich relative sunspot
numbers\footnote{http://www.ngdc.noaa.gov/stp/spaceweather.html}
($R_{\mathrm{z}}$) of the more reliable data since Cycle 8 (up to
November 2010). Some parameters of the solar cycle are listed in
Table~\ref{Tab:1}, where $R_{\rm max}$ and $R_{\rm min}$ are the
maximum and minimum amplitudes of a solar cycle, respectively.
Here $\beta_{\rm a}$ is the rising rate at $\Delta m$ months
entering into the cycle,
\begin{equation}
  \label{Eq:rates}
   \beta_{\rm a}=\Delta R_{\rm z}/\Delta m,
\end{equation}
where $\Delta R_{\rm z}=R_{\rm z}(\Delta m)-R_{\rm min}$ is the
increment of $R_{\rm z}$ from $R_{\rm min}$ in the time interval
of $\Delta m$ (months). At the present time, $\Delta m=24$
(months) for the data available in the current cycle (24). $S$ is
the skewness of the cycle, a measure of symmetry about its
maximum,
  \begin{equation}
  \label{Eq:skewness}
   S=\frac{\sum_{i=1}^N(y_i-\overline{y})^3}{(N-1)\sigma^3},
\end{equation}
where $y=R_{\rm z}$ is the data series for a solar cycle,
$\overline{y}$ is the mean, $N$ the number of data points, and
$\sigma$ the standard deviation. Positive (negative) skewness
indicates that the distribution is skewed to the right (left),
with a longer tail to the right (left) of the maximum.

\begin{table}[h!!!]
\small 
 \centering
\begin{minipage}[]{120mm}
  \caption[]{ Parameters since cycle $n=8$.
 }\label{Tab:1}\end{minipage}
\tabcolsep 2.0mm  
 \begin{tabular}{rrrrrr}
  \hline\noalign{\smallskip}
$n$ &  $R_{\rm min}$ &$\beta_{\rm a}(21)$ &$\beta_{\rm
a}(24)^{\mathrm{e}}$ &$R_{\rm max}$ & $S$ \\
  \hline\noalign{\smallskip}
 8  & 7.3  &2.85  &3.31  &146.9  &0.28  \\
 9  &10.6  &1.10  &1.25  &131.9  &0.51  \\
10  & 3.2  &1.24  &1.37  & 98.0  &0.14  \\
11  & 5.2  &2.48  &2.63  &140.3  &0.52  \\
12  & 2.2  &1.65  &1.73  & 74.6  &0.09  \\
13  & 5.0  &2.32  &2.38  & 87.9  &0.35  \\
14  & 2.7  &1.28  &1.37  & 64.2 &$-0.09$\\
15  & 1.5  &1.94  &1.95  &105.4  &0.33  \\
16  & 5.6  &1.29  &1.73  & 78.1  &0.02  \\
17  & 3.5  &1.46  &1.79  &119.2  &0.12  \\
18  & 7.7  &2.11  &2.47  &151.8  &0.16  \\
19  & 3.4  &4.07  &4.80  &201.3  &0.30  \\
20  & 9.6  &1.94  &2.42  &110.6  &0.03  \\
21  &12.2  &2.32  &2.68  &164.5  &0.13  \\
22  &12.3  &3.88  &4.54  &158.5  &0.16  \\
23  & 8.0  &1.95  &2.14  &120.8  &0.30  \\
24  & 1.7  &0.75  &1.03\\
\hline
$\overline{x}^{\mathrm{a}}$         &6.3    &2.12 &2.04 &122.1 &0.21   \\
$\sigma_{\rm x}^{\mathrm{b}}$       &3.5    &0.88 &0.91 &37.4  &0.17   \\
$r^{\mathrm{c}}$                    &0.39   &0.36 &0.15 &0.33  &$-0.32$\\
CL(\%)$^{\mathrm{d}}$               &86.4   &82.7 &$<50$&78.4  &76.1   \\
  \noalign{\smallskip}\hline
\end{tabular}
 \begin{list}{}{}
  \item [$ ^{\mathrm{a}}$] Average for $n=8$-23.
  \item [$ ^{\mathrm{b}}$] Standard deviation.
  \item [$ ^{\mathrm{c}}$] Correlation coefficient of temporal trend.
  \item [$ ^{\mathrm{d}}$] The confidence level (\%).
  \item [$ ^{\mathrm{e}}$] At the current state.
  \end{list}
\end{table}

   \begin{figure}[h!!!]
   \centering
   \includegraphics[width=3.7cm, angle=0]{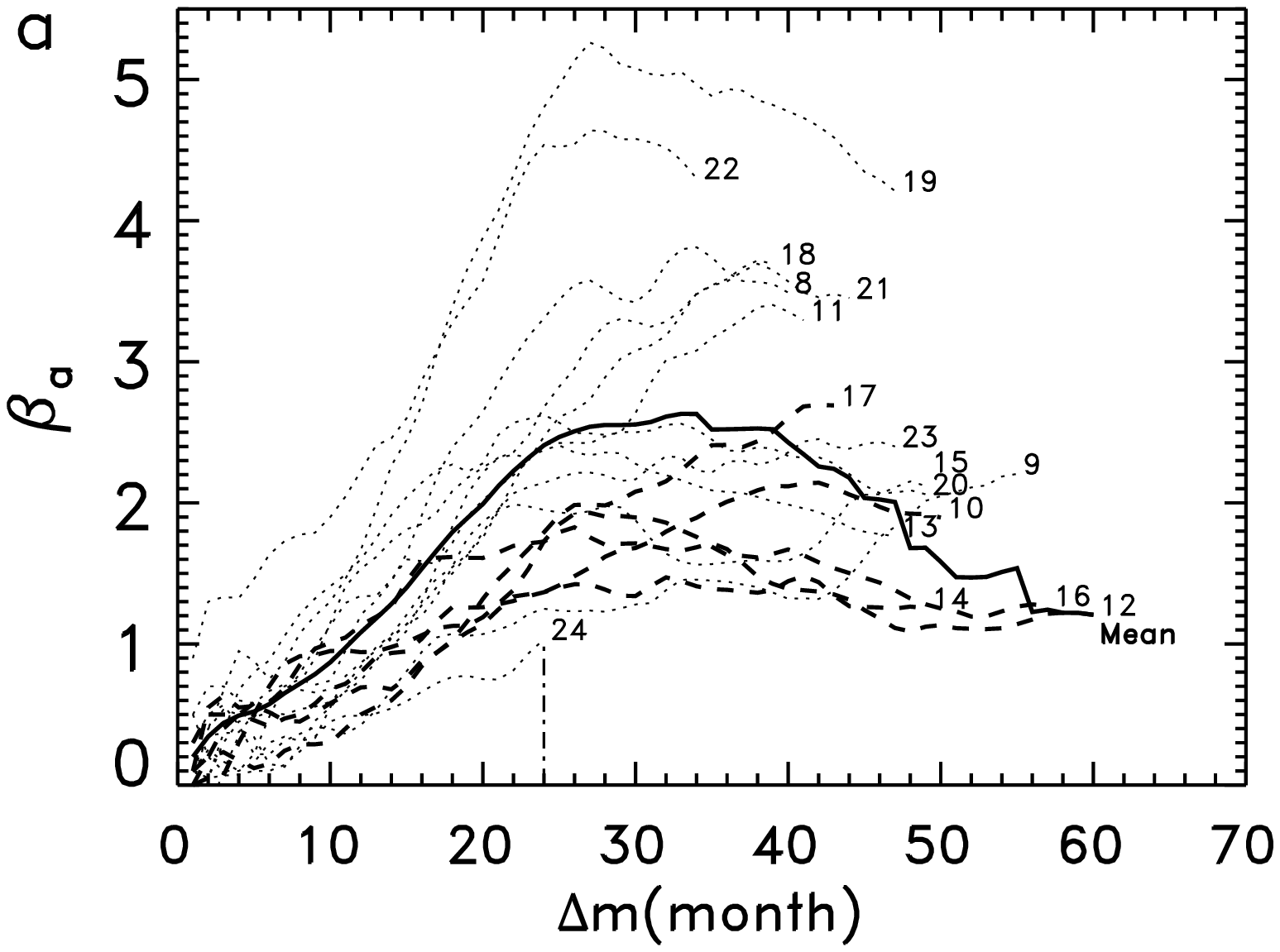}
   \includegraphics[width=3.7cm, angle=0]{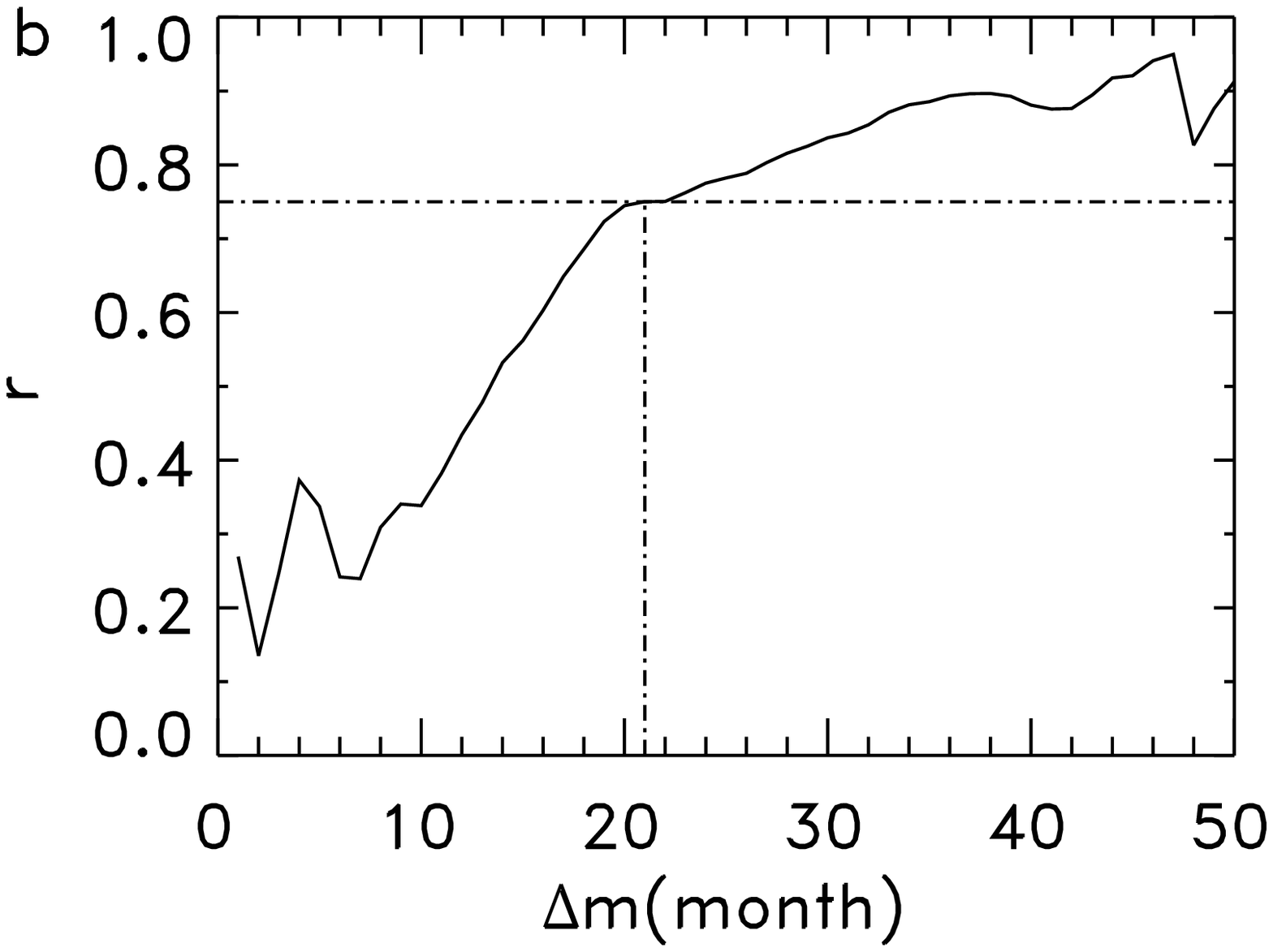}
   \caption{ (a) $\beta_{\rm a}$ as a function of $\Delta m$
   for Cycles 8-24. The dashed lines indicate $\beta_{\rm a}$
   for Cycles 14, 10, 17, 16 and 12. The thick solid line
   indicates the averages for the data available.
   (b) Correlation coefficient ($r$) between $R_{\rm max}$ and $\beta_{\rm a}$.
   }
   \label{Fig:1}
   \end{figure}

In Table~\ref{Tab:1}, $\overline{x}$ and $\sigma$ are the average
and standard deviation of the parameters for Cycles $n=8$-23,
respectively, $r$ is the correlation coefficient of the parameter
with time (temporal variation), and CL is the confidence level.

It is shown in Table~\ref{Tab:1} that $R_{\rm min}$, $\beta_{\rm
a}$ and $R_{\rm max}$ all show an increasing trend with time
($r>0$) at the confidence level (CL) around 80\%, and that $S$
used to be positive ($\overline{S}=0.21$), meaning that the
decline time tends to be longer than the rise time for most solar
cycles. However, this asymmetry seems to decrease as can be seen
from the decreasing temporal trend ($r=-0.32$) in $S$.

The $\beta_{\rm a}$ values for Cycles 8-24 are shown in
Fig.~\ref{Fig:1}(a). It is seen in this figure that $\beta_{\rm
a}$ varies approximately linearly with $\Delta m$ at the early
phase of the cycle (about $\Delta m\le26$) and has decreases since
then. The correlation coefficient ($r$) between $R_{\rm max}$ and
$\beta_{\rm a}$ varies with an increasing trend
(Fig.~\ref{Fig:1}b) and $r>0.75$ when $\Delta m\ge21$. First, we
take $\Delta m=21$ (months) as an example to describe the method
of similar cycles. Then, this method is applied to $\Delta
m=18,19,\cdots,24$ (months) for predicting Cycle 24.

\section{Correlation between some parameters } \label{sec:Correlation}

\begin{figure}[h!!!]
   \centering
   \includegraphics[width=8.0cm, angle=0]{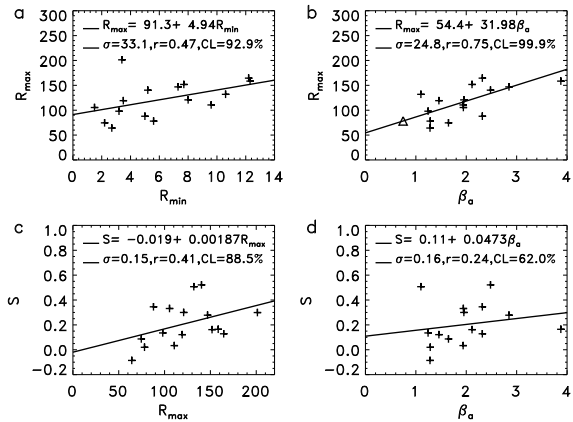}
   \caption{ Scatter plots of (a) $R_{\mathrm{max}}$ against $R_{\mathrm{min}}$,
   (b) $R_{\mathrm{max}}$ against $\beta_{\rm a}$,
   (c) $S$ against $R_{\mathrm{max}}$,
   (d) $S$ against $\beta_{\rm a}$.
   }
   \label{Fig:2}
   \end{figure}

Of the  parameters describing the solar cycle, only
$R_{\mathrm{min}}$ and $\beta_{\rm a}$ are known before the timing
of $R_{\mathrm{max}}$. Therefore, both $R_{\mathrm{min}}$ and
$\beta_{\rm a}$ can be taken as indicators for the subsequent
$R_{\mathrm{max}}$.  Figure\,\ref{Fig:2}(a) shows the scatter plot
of $R_{\mathrm{max}}$ against $R_{\mathrm{min}}$ with the
least-square-fit linear regression equation given by
\begin{equation}
  \label{Eq:RmaxRmin}
   R_{\mathrm{max}} = 91.3 +4.94 R_{\mathrm{min}},
   \sigma=33.1,
\end{equation}
where $\sigma$ is the standard deviation. It is well known that
$R_{\mathrm{max}}$ is weakly correlated with the preceding
$R_{\mathrm{min}}$ ($r_{\rm R}=0.47$), at the 92.9\% confidence
level, so that a small $R_{\mathrm{min}}$ tends to be followed by
a weak cycle \citep{Hathaway02,Li09}. However, since this
correlation is weak, one can hardly obtain an accurate prediction
of $R_{\mathrm{max}}$ by $R_{\mathrm{min}}$ alone \citep{Du210}.

Figure\,\ref{Fig:2}(b) shows the scatter plot of
$R_{\mathrm{max}}$ against $\beta_{\mathrm{a}}$. The best linear
relationship between them is
\begin{equation}
  \label{Eq:Rmaxbeta}
   R_{\mathrm{max}} = 54.4 +31.98 \beta_{\mathrm{a}},
   \sigma=24.8.
\end{equation}
One can see that $R_{\mathrm{max}}$ is well correlated with the
rising rate $\beta_{\rm a}$ ($r_{\rm \beta}=0.75$) at the 99.9\%
confidence level. Therefore, $\beta_{\rm a}$ is a good predictor
for the subsequent $R_{\mathrm{max}}$. Substituting the value of
$\beta_{\mathrm{a}}$ (0.75) for $\Delta m=21$ into this equation,
the peak size of Cycle 24 can be estimated to be
$R_{\mathrm{max}}(24)= 78.3\pm24.8$ (triangle).

The scatter plots of $S$ against $R_{\mathrm{max}}$ and $S$
against $\beta_{\rm a}$ are shown in Figs.~\ref{Fig:2}(c) and (d),
respectively. The correlation coefficients involved are listed in
Table~\ref{Tab:2}.

\begin{table}[h!!!]
\small \centering
\begin{minipage}[]{120mm}
  \caption[]{ Cross-correlation Coefficients ($r$).
 }\label{Tab:2}\end{minipage}
\tabcolsep 4.5mm  
 \begin{tabular}{ll|cc}
  \hline\noalign{\smallskip}
 $y$     & $x$             & $r$  & CL  \\%
  \hline\noalign{\smallskip}
$R_{\rm max}$  &$R_{\rm min}$    &0.47    &92.9\%         \\
$R_{\rm max}$  &$\beta_{\rm a}$  &0.75    &99.9\%         \\
$S$            &$R_{\rm max}$    &0.41    &88.5\%  \\
$S$            &$\beta_{\rm a}$  &0.24    &62.0\%         \\
$S$            &$R_{\rm min}$    &0.10         \\
  \noalign{\smallskip}\hline
\end{tabular}
\end{table}

\section{Similar cycles } \label{sec:Similar}

It should be pointed out in Fig.~\ref{Fig:2}(c) that $S$ is
positively correlated with $R_{\mathrm{max}}$ ($r=0.41$), so that
similar cycles (with approximatively the same $R_{\mathrm{max}}$)
tend to have similar shapes, which is a key point in the concept
of similar cycles although $r$ is not high. Because if $S$ were
uncorrelated with $R_{\mathrm{max}}$, the concept of similar
cycles would not have been used any longer. However, as
$R_{\mathrm{max}}$ is unknown in advance, the shape of an upcoming
cycle cannot be described directly from the measurement of
$R_{\mathrm{max}}$, which is usually estimated by an appropriation
technique.

Because $R_{\mathrm{max}}$ has a much higher correlation
coefficient with $\beta_{\rm a}$ ($r=0.75$) than with
$R_{\mathrm{min}}$ ($r=0.47$) and the correlation coefficient of
$S$ with $\beta_{\rm a}$ ($r=0.24$) is slightly higher than that
of $S$ with $R_{\mathrm{min}}$ ($r=0.10$), $\beta_{\rm a}$ is a
much better predictor for $R_{\mathrm{max}}$ than
$R_{\mathrm{min}}$. However, the two predictors of $\beta_{\rm a}$
and $R_{\mathrm{min}}$ are expected to provide useful information
(size and asymmetry) on the subsequent cycle, because both of them
can be directly measured a few months after the minimum.

\subsection{Degree of similarity} \label{subsec:similarity}

\begin{figure}[h!!!]
   \centering
   \includegraphics[width=8.0cm, angle=0]{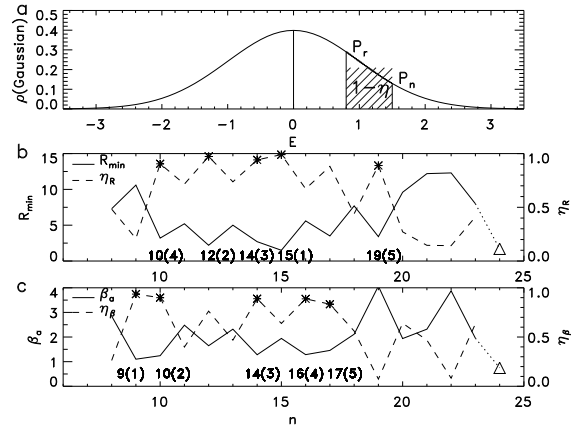}
   \caption{ (a) Gaussian distribution.
     (b) $R_{\mathrm{min}}$ (solid) and $\eta_{\mathrm{R}}$ (dashed).
     (c) $\beta_{\mathrm{a}}$ (solid) and $\eta_{\mathrm{\beta}}$ (dashed).
     The numbers indicate the similar cycles and their orders in brackets.}
   \label{Fig:3}
   \end{figure}

For a parameter $x=\{x_i,i=1,2,\cdots,m\}$, the probability
density of the normalized deviation,
\begin{equation}
  \label{Eq:E0}
   E_i = \frac{\Delta x_i}{\sigma_{\rm{x}}}= \frac{x_i-\overline{x}}{\sigma_{\rm{x}}},
\end{equation}
approximately satisfies a normal distribution (Fig.~\ref{Fig:3}a),
\begin{equation}
  \label{Eq:normal}
   \rho(E_i) = \frac{1}{\sqrt{2\pi}} e^{-E_i^2/2},
\end{equation}
where $\overline{x}$ and $\sigma_{\rm{x}}$ are the mean and
standard deviation, respectively (Table~\ref{Tab:1}). The
probability that $\Delta x_i$ falls in the range [$\Delta
x_{\rm{r}}$, $\Delta x_{\rm{n}}$] is
\begin{equation}
  \label{Eq:difP}
   \Delta P (n)= \frac{1}{\sqrt{2\pi}}\int_{E_{\rm{r}}}^{E_{\rm{n}}} e^{-t^2/2} dt,
\end{equation}
where $x_{\rm{r}}$ is a referenced value (a predictor for Cycle
$n_{\rm{r}}=24$) and $x_{\rm{n}}$ the value in the past cycle
($n$).

A small value of $|\Delta P(n)|$ indicates that
$x_{\rm{n}}(=\overline{x}+\Delta x_{\rm{n}})$ is close to
$x_{\rm{r}}(=\overline{x}+\Delta x_{\rm{r}})$, in which case Cycle
$n$ is called a similar cycle of $n_{\rm{r}}$ around $x_{\rm{r}}$.
The smaller the $|\Delta P(n)|$ is, the more similar the two
parameters are. Because $0\le|\Delta P(n)|\le1$ for either
$x_{\rm{r}}\ge x_{\rm{n}}$ or $x_{\rm{r}}< x_{\rm{n}}$, we define
$1-|\Delta P(n)|$ as a measure to describe the ``degree of
similarity" around $x_{\rm{r}}$,
\begin{equation}
  \label{Eq:similarity}
   \eta(n) = 1-\frac{1}{\sqrt{2\pi}}\left|\int_{E_{\rm{r}}}^{E_{\rm{n}}} e^{-t^2/2}
   dt\right|.
\end{equation}
If $\eta (n)=1$, the two parameters are identical,
$x_{\rm{n}}=x_{\rm{r}}$ (100\% similarity); if $\eta (n)=0$, there
is no similarity between them. In application, some (five in this
study) of the largest values of $\eta(n)$ are selected and their
cycles are called the similar cycles of $n_{\rm{r}}$ around
$x_{\rm{r}}$.

\subsection{Similar cycles used in predicting $R_{\mathrm{max}}$} \label{subsec:Rm}

Figure\,\ref{Fig:3}(b) shows the values of $R_{\mathrm{min}}$
(solid) for Cycles $n=8$-23, the referenced one
($R_{\mathrm{min}}(24)=1.7$, triangle) for Cycle  $n_{\rm{r}}=
24$, and the values of $\eta$ ($\eta_{\mathrm{R}}$, dashed) around
$R_{\mathrm{min}}(24)$. The five biggest values of
$\eta_{\mathrm{R}}$ (asterisks) are in Cycles $n_{\mathrm{R}}=15,
12, 14, 10$ and 19 (in that order), which are called the similar
cycles of Cycle 24 around $R_{\mathrm{min}}$.

Figure\,3(c) shows the values of $\beta_{\mathrm{a}}$ (solid), the
referenced one ($\beta_{\mathrm{a}}(24)=0.75$, triangle), and the
values of $\eta$ ($\eta_{\mathrm{\beta}}$, dashed) around
$\beta_{\mathrm{a}}(24)$. The five biggest values of
$\eta_{\mathrm{\beta}}$ (asterisks) are in Cycles
$n_{\mathrm{\beta}}=9, 10, 14, 16$ and 17 (in that order), which
are called the similar cycles of Cycle 24 around
$\beta_{\mathrm{a}}$. These similar cycles ($n_{\mathrm{\beta}}$)
are partly different from the above ones around $R_{\mathrm{min}}$
($n_{\mathrm{R}}$). Since the correlation coefficient of
$R_{\mathrm{max}}$ with $\beta_{\mathrm{a}}$
($r_{\rm{\beta}}=0.75$) is much higher than that
($r_{\rm{R}}=0.47$) of $R_{\mathrm{max}}$ with $R_{\mathrm{min}}$,
$n_{\mathrm{\beta}}$ should be more reliable than $n_{\mathrm{R}}$
for describing the information of Cycle $n_{\mathrm{r}}$.

From $\eta_{\rm{R}}$ and $\eta_{\mathrm{\beta}}$, a synthesis
degree of similarity around both $R_{\mathrm{min}}$ and
$\beta_{\mathrm{a}}$ can be defined as
\begin{equation}
  \label{Eq:eta1}
   \eta_{\rm{s}}(n) = \frac{\eta_{\mathrm{R}}(n)r_{\rm{R}}^2+\eta_{\mathrm{\beta}}(n)r_{\rm{\beta}}^2}{r_{\rm{R}}^2+r_{\rm{\beta}}^2},
\end{equation}
here the weights are taken as the coefficients of determination,
$W_{\rm{R}}=r^2_{\rm{R}}$ and $W_{\rm{\beta}}=r^2_{\rm{\beta}}$
for $R_{\mathrm{min}}$ and $\beta_{\mathrm{a}}$, respectively. The
reason for such a selection is that about $r^2_{\rm{R}}$
($r^2_{\rm{\beta}}$) of the variations in $R_{\mathrm{max}}$ can
be explained by the correlation of $R_{\mathrm{max}}$ with
$R_{\mathrm{min}}$ ($\beta_{\mathrm{a}}$).

   \begin{figure}[h!!!]
   \centering
   \includegraphics[width=8.0cm, angle=0]{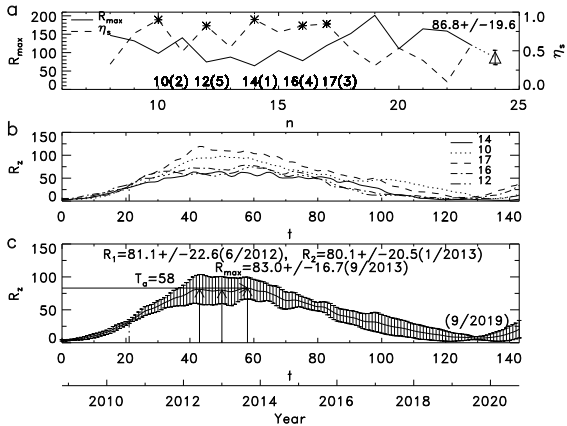}
   \caption{ (a) $R_{\mathrm{max}}$ (solid) and $\eta_{\rm{s}}$ (dashed).
   (b) The values of $R_{\mathrm{z}}$ for similar cycles $n_s=14, 10, 17, 16$ and 12.
   (c) The predicted monthly values of $R_{\mathrm{z}}$ for Cycle 24 with error bars.
   }
   \label{Fig:4}
   \end{figure}

Figure\,\ref{Fig:4}(a) shows the values of $R_{\mathrm{max}}$
(solid) and $\eta_{\rm{s}}$ (dashed). The five biggest values of
$\eta_{\rm{s}}$ (asterisks) are in Cycles $n_{\rm{s}}=14, 10, 17,
16$ and 12 (in that order), which are called the similar cycles of
Cycle 24 around both $R_{\mathrm{min}}$ and $\beta_{\mathrm{a}}$.
From these values, the peak size of Cycle 24 can be predicted as
the weighted-average of those of the similar cycles,
\begin{equation}
  \label{Eq:Rmax24}
  \begin{array}{rcl}
   R_{\mathrm{max}}(24) &=& \frac{ \sum_{i=1}^{5}R_{\mathrm{max}}(n_{\rm{s}}(i))W(i) }{ \sum_{i=1}^{5}W(i)
   }, \\
   W(i)  &=& \eta_{\rm{s}}(n_{\rm{s}}(i)),\\
  \end{array}
\end{equation}
where the weights $W(i)$ are taken as the (synthesis) degrees of
similarity $\eta_{\rm{s}}(n_{\rm{s}}(i))$. The more similar a
cycle is to $n_{\rm{r}}$, the more its weight it should be
included. The standard deviation is correspondingly defined as
\begin{equation}
  \label{Eq:sig24}
   \sigma_{\mathrm{max}}(24) = \sqrt{ \frac{ \sum_{i=1}^{5} [R_{\mathrm{max}}(n_{\rm{s}}(i))- R_{\mathrm{max}}(24)]^2W(i) }{
    \sum_{i=1}^{5}W(i) }
   }.
\end{equation}
According to the above equations, the peak size of Cycle 24 is
predicted to be $R_{\mathrm{max}}(24)=86.8\pm 19.6$ (triangle).

\subsection{Similar cycles used in predicting monthly values} \label{subsec:shape}

Now, we use the above technique to predict the monthly values
(shape) of Cycle 24. Figure\,\ref{Fig:4}(b) shows the monthly
values of $R_{\mathrm{z}}$ for the similar cycles ($n_{\rm{s}}$)
from the starting points of the cycles. The $R_{\mathrm{z}}$ value
of the $t^{th}$ month in Cycle 24 is predicted as the
weighted-average of the corresponding $R_{\mathrm{z}}$ values for
the same month in the similar cycles,
\begin{equation}
  \label{Eq:Rz}
   \overline{R}_{\mathrm{z}}(t) = \frac{ \sum_{i=1}^{5}R_{\mathrm{z}}(t,n_{\rm{s}}(i))W(i) }{ \sum_{i=1}^{5}W(i)
   }.
\end{equation}
Its standard deviation is
\begin{equation}
  \label{Eq:sig}
   \sigma_{\mathrm{z}}(t) = \sqrt{ \frac{ \sum_{i=1}^{5} (R_{\mathrm{z}}(t,n_{\rm{s}}(i))- \overline{R}_{\mathrm{z}}(t))^2W(i) }{
    \sum_{i=1}^{5}W(i) }
   }.
\end{equation}

The results since November 2008 are shown in Fig.~\ref{Fig:4}(c).
From these results, we can obtain the maximum value (83.0), the
standard deviation (16.7), the rise time ($T_{\mathrm{a}}=58$
months), and the cycle length (130 months). Cycle 24 is therefore
predicted to peak around September 2013 with a size of about
$R_{\mathrm{max}}=83.0\pm 16.7$, and to end around September 2019.
This size is near that (86.8) in section \ref{subsec:Rm} and that
(78.3) in Section \ref{sec:Correlation}. It should be pointed out
in Fig.~\ref{Fig:4}(c) that the shape near the peak is rather
flat. Besides the highest peak (83.0) in September 2013, there are
two weak peaks preceding it: the first is around June 2012 (81.1)
and the second is around January 2013 (80.1). If this result is
true, it implies that Cycle 24 will have multiple peaks with the
last peak being higher. On average, Cycle 24 may probably peak
around January 2013 $\pm8$ (months).

\subsection{Prediction results at $\Delta m$ = 18-24 months} \label{subsec:m}

\begin{figure}[h!!!]
   \centering
   \includegraphics[width=8.0cm, angle=0]{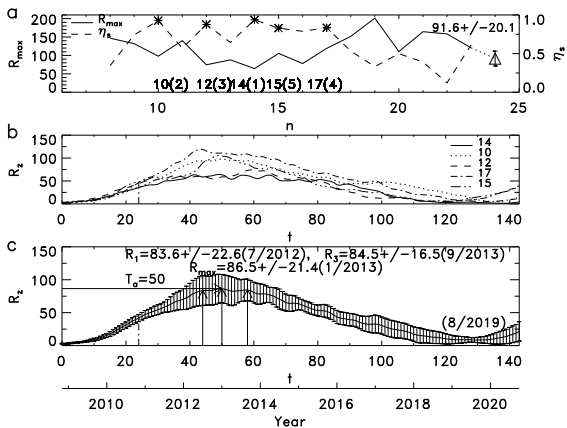}
   \caption{ Similar to Fig.~\ref{Fig:4} but using the $\beta_{\mathrm{a}}$ values at $\Delta m=24$ (month).
   (a) $R_{\mathrm{max}}$ (solid) and $\eta_{\rm{s}}$ (dashed).
   (b) The values of $R_{\mathrm{z}}$ for similar cycles $n_{\rm{s}}=14, 10, 12, 17$ and 15.
   (c) The predicted monthly values of $R_{\mathrm{z}}$ for Cycle 24 with error bars.
   }
   \label{Fig:5}
   \end{figure}

Since $\beta_{\mathrm{a}}$ is a temporal variable of $\Delta m$,
the result derived from $\beta_{\mathrm{a}}$ may depend on $\Delta
m$, similar to the prediction by a simple function to describe the
shape of the solar cycle \citep{Hathaway94,Du11b}. In this
section, we apply the above technique to the current state
($\Delta m=24$), as shown in Fig.~\ref{Fig:5}. The similar cycles
at $\Delta m=24$ ($n_{\rm{s}}=$14, 10, 12, 17 and 15) are slightly
different from those at $\Delta m=21$ ($n_{\rm{s}}=$14, 10, 17, 16
and 12) in Fig.~\ref{Fig:4}. In Fig.~\ref{Fig:5}(a), the peak size
of Cycle 24 is predicted to be $R_{\mathrm{max}}(24)=91.6\pm 20.1$
(triangle) based on the $R_{\mathrm{max}}$ values of the similar
cycles, slightly higher than that ($R_{\mathrm{max}}(24)=86.8\pm
19.6$) in Fig.~\ref{Fig:4}(a).

In Fig.~\ref{Fig:5}(c), the highest peak is
$R_{\mathrm{max}}=86.5\pm21.4$ in January 2013 (from the predicted
monthly $R_{\mathrm{z}}$ in Cycle 24), which is near the second
peak in Fig.~\ref{Fig:4}(c). In addition, there are two shoulders:
one in July 2012 (83.8) and another in September 2013 (84.5). The
former is near the first peak (June 2012) and the latter is just
the third peak (September 2013) in Fig.~\ref{Fig:4}(c).

\begin{table*}[!tb]
\small 
 \centering
\begin{minipage}[]{120mm}
  \caption[]{ Prediction Results for $\Delta m=18,19,\cdots,24$ (months).
 }\label{Tab:3}\end{minipage}
\tabcolsep 2.0mm  
 \begin{tabular}{ccccccc}
  \hline\noalign{\smallskip}
$\Delta m$  &$\beta_{\rm a}$  &$n_{\rm s}$ &$R_{\rm max}$(1st
peak)
&$R_{\rm max}$(2nd peak) &$R_{\rm max}$(3rd peak) &ending\\%
  \hline\noalign{\smallskip}
18  &0.77  &10,17,14,16,15 &85.9$\pm$20.5(6/2012) &89.4$\pm$19.2(12/2012)&86.4$\pm$15.8(8/2013) &2/2019 \\
19  &0.77  &10,17,14,16,12 &81.8$\pm$22.7(6/2012) &80.7$\pm$20.6(1/2013) &83.5$\pm$16.8(9/2013)  &9/2019 \\
20  &0.75  &10,14,17,16,12 &81.4$\pm$22.6(6/2012) &80.4$\pm$20.6(1/2013) &83.2$\pm$16.7(9/2013) &9/2019 \\
21  &0.75  &14,10,17,16,12 &81.1$\pm$22.6(6/2012) &80.1$\pm$20.5(1/2013) &83.0$\pm$16.7(9/2013) &9/2019 \\
22  &0.81  &14,10,17,12,16 &81.0$\pm$22.5(6/2012) &80.0$\pm$20.5(1/2013) &82.9$\pm$16.7(9/2013) &9/2019 \\
23  &0.93  &14,10,12,17,16 &80.9$\pm$22.5(6/2012) &79.9$\pm$20.5(1/2013) &82.8$\pm$16.7(9/2013) &9/2019 \\
24  &1.03  &14,10,12,17,15 &83.6$\pm$22.6(7/2012) &86.5$\pm$21.4(1/2013) &84.5$\pm$16.5(9/2013) &8/2019 \\
\hline
$\overline{x}$ &   &       &82.2$\pm$22.3(6/2012) &82.4$\pm$20.5(1/2013) &83.8$\pm$16.6(9/2013)  &9/2019 \\%
  \noalign{\smallskip}\hline
\end{tabular}
\end{table*}

Using the same technique, we tested the predictive power of this
method at $\Delta m$ = 18\,-\,24 month. The results are listed in
Table~\ref{Tab:3}, where the third column indicates the similar
cycles for a given $\Delta m$ (first column); the fourth\,-\,sixth
columns correspond to the first\,-\,third peaks, respectively; and
the last column is the ending time of Cycle 24. The last row shows
the relevant averages. It is seen in this table that there are not
significant differences in the results as the cycle progresses
although the similar cycles may have a small difference ($n_{\rm
s}$) for different $\Delta m$. The three peaks for different
$\Delta m$ are near each other either in size or in date. If the
maximum average size and the middle date is taken as those for the
peak of the cycle, Cycle 24 is predicted to peak around January
2013 $\pm$ 8 (month) with a size of about $R_{\rm max}=84\pm17$.

\section{Discussions and Conclusions}
\label{sec:Discussions}

A concept, the degree of similarity ($\eta$), is proposed to
quantitatively describe the similarity of a parameter of a solar
cycle relative to a referenced one, and the prediction method of
similar cycles is further developed. The degrees of similarity are
used as the weights in the weighted-average of the values of a
parameter in the similar cycles so as to obtain a predicted one in
the referenced cycle, where we have considered the fact that more
weights should be paid to the cycles that are more similar to the
referenced one.

In this study, we used two predictors, the preceding minimum
($R_{\mathrm{min}}$) and rising rate ($\beta_{\mathrm{a}}$), to
define a synthesis degree of similarity ($\eta_{\rm s}$) by
averaging the corresponding $\eta$ values with the weights given
by the coefficients of determination of $R_{\mathrm{max}}$ with
$R_{\mathrm{min}}$ and $\beta_{\rm a}$, respectively. From this
method, Cycle 24 is predicted to peak around January 2013 $\pm$ 8
(months) with a size of about $R_{\rm max}=84\pm17$ and to end
around September 2019. This result is slightly lower than that
($100.2\pm7.5$) predicted by \cite{WangJL09} based on their
similar-cycle method. It is near $80\pm30$ \citep{Schatten05},
$\sim80$ \citep{Kitiashvili08}, and $\sim 85$ \citep{Jiang07}
based on polar field or solar dynamo models, but much lower than
$\sim 167$ \citep{Dikpati06} based on a modified flux-transport
dynamo model.

It should be pointed out from Section~\ref{subsec:Rm} that (i)
considering a single parameter $R_{\mathrm{min}}$, the most
similar cycle (to Cycle 24) is $n=15$ since
$R_{\mathrm{min}}(15)=1.5$ is close to $R_{\mathrm{min}}(24)=1.7$;
(ii) considering another single parameter $\beta_{\mathrm{a}}$,
the most similar cycle is $n=9$ as $\beta_{\mathrm{a}}(9)=1.10$ is
close to $\beta_{\mathrm{a}}(24)=0.75$; and (iii) considering both
parameters $R_{\mathrm{min}}$ and $\beta_{\mathrm{a}}$, the most
similar cycle is $n=14$. In this study, the similar cycles are
selected from two parameters ($R_{\mathrm{min}}$,
$\beta_{\mathrm{a}}$) to avoid an accidental error caused by a
single parameter. In terms of the synthesis degree of similarity
($\eta_{\rm s}$), Cycle 15 is not a similar cycle since
$\beta_{\mathrm{a}}(15)=1.94$ is not so near to
$\beta_{\mathrm{a}}(24)$, although $R_{\mathrm{min}}(15)$ is close
to $R_{\mathrm{min}}(24)$, and Cycle 9 is not a similar cycle as
$R_{\mathrm{min}}(9)=10.6$ is far from $R_{\mathrm{min}}(24)$
although $\beta_{\mathrm{a}}(9)$ is close to
$\beta_{\mathrm{a}}(24)$. Therefore, the most similar cycle
($n=14$) is that with maximum $\eta_{\rm s}$ which is the
synthesis effect of both $R_{\mathrm{min}}(14)=2.7$ and
$\beta_{\mathrm{a}}(14)=1.28$, although Cycle 14 is not the most
similar cycle with either $R_{\mathrm{min}}$ or
$\beta_{\mathrm{a}}$ alone. Because $\beta_{\mathrm{a}}$ varies
with the progression ($\Delta m$) of the cycle, the results from
both $R_{\mathrm{min}}$ and $\beta_{\mathrm{a}}$ may have a slight
differences (Table~\ref{Tab:3}).

Near the time of a solar minimum, geomagnetic activity  is a much
better indicator for the ensuing maximum amplitude
($R_{\mathrm{max}}$) of the solar cycle~\citep{Ohl66} than the
solar minimum ($R_{\mathrm{min}}$). This study shows that the
rising rate ($\beta_{\mathrm{a}}$) at the early phase of a solar
cycle is also a good indicator for the subsequent
$R_{\mathrm{max}}$. This parameter has the advantage that it only
needs the $R_{\mathrm{z}}$ series itself. It reflects the initial
physical process of solar magnetic activities.

Similar cycles are usually defined as those whose values of
$R_{\mathrm{max}}$ (or other parameters) satisfy a given condition
\citep{Gleissberg71,WangJL97,WangJL02a,WangJL09,Miao08},
\begin{equation}
  \label{Eq:detRm}
   |R_{\mathrm{max}}-R_{\mathrm{max}}(n_{\rm r})| \leq \Delta,
\end{equation}
where $R_{\mathrm{max}}(n_{\rm r})$ is the referenced value and
$\Delta$ is a given limit ({\it e.g.,} 10 or 20). By averaging the
values of a parameter (rise time, decline time or monthly
$R_{\mathrm{z}}$ and so on) in these cycles, the corresponding one
in the predicted cycle could be obtained. This technique has
considered the cycles that have similar $R_{\mathrm{max}}$ to
$R_{\mathrm{max}}(n_{\rm r})$. However, how close these cycles are
to the referenced one has not been considered, as a simple
averaging method ($W(i)=1$) was used in this technique. Besides,
$R_{\mathrm{max}}(n_{\rm r})$ is usually unknown, so it also needs
to be predicted. The error in $R_{\mathrm{max}}(n_{\rm r})$ may
probably propagate into the other parameters that are derived from
it. In contrast, this study used the two directly measured
parameters ($R_{\mathrm{min}}$, $\beta_{\rm{a}}$) to derive all
the information of a cycle to be predicted (24).

The central idea of a similar cycle is that solar cycles with
approximatively the same $R_{\mathrm{max}}$ tend to have similar
shapes, which is empirical and has not been studied physically as
far as we know. It is shown in Fig.~\ref{Fig:2}(c) that $S$ is
weakly correlated with $R_{\mathrm{max}}$ ($r=0.41$), which is a
key point in the concept of similar cycles. Under this
relationship, we can use the concept that solar cycles with
approximate $R_{\mathrm{max}}$ might have similar rise times,
cycle lengths and cycle shapes, so we can estimate it in the
referenced cycle by averaging the corresponding values in similar
cycles. However, since $R_{\mathrm{max}}$ is unknown in advance,
the shape of an upcoming cycle cannot be estimated directly from
$R_{\mathrm{max}}$. In this study, we used both the rising rate
($\beta_{\mathrm{a}}$) and solar minimum ($R_{\mathrm{min}}$) to
select the similar cycles, which is based on the correlations of
$R_{\mathrm{max}}$ with both $\beta_{\mathrm{a}}$
($r_{\mathrm{\beta}}\sim0.75$) and $R_{\mathrm{min}}$
($r_{\mathrm{R}}=0.47$). If one uses other parameters (e.g., the
preceding decline time) to find some similar cycles, the
correlations between $R_{\mathrm{max}}$ and these parameters are
also needed (even if they are not strong).

One advantage of the similar-cycle method is that it does not
involve the details of a physical process. The actual physical
process may be rather complicated; its dynamical mechanism is not
very clear at present and may not be described by a simple linear
or non-linear relationship. Whatever the process is, a similar
process may likely occur if the levels of activity are similar,
which is what the similar-cycle method can and wants to do. This
is similar to treating the complex process as a piecewise
function.

The main points in this study can be summarized as follows.
\begin{enumerate}
\item A concept, the degree of similarity ($\eta$), is proposed to
quantitatively describe the similarity about a parameter for a
solar cycle relative to a referenced one.
 \item For two parameters, the preceding minimum ($R_{\mathrm{min}}$) and
rising rate ($\beta_{\mathrm{a}}$), a synthesis degree of
similarity ($\eta_{\rm s}$) is defined as the weighted-average of
the corresponding ones with the weights given by the coefficients
of determination of $R_{\mathrm{max}}$ with $R_{\mathrm{min}}$ and
$\beta_{\rm a}$, respectively.
 \item The prediction method of similar cycles is further
developed with the weights given by the (synthesis) degrees of
similarity.
 \item  As an application of this method, Cycle 24 is predicted to peak around 
 January 2013 $\pm$ 8 (months) with a size of about $R_{\mathrm{max}}=84\pm 17$ and to end
around September 2019.
\end{enumerate}

\section*{Acknowledgments}
This work is supported by the National Natural Science Foundation
of China (NSFC) through grants 10973020, 40890161 and 10921303,
and the National Basic Research Program of China through grant No.
2011CB811406.


\end{document}